\documentclass[fleqn,12pt,twoside]{article}
\usepackage[headings]{espcrc1}

\readRCS
$Id: espcrc1.tex,v 1.2 2004/02/24 11:22:11 spepping Exp $
\usepackage{graphicx,psfig,epsfig}

\newcommand{\AmS}{{\protect\the\textfont2
  A\kern-.1667em\lower.5ex\hbox{M}\kern-.125emS}}
\def\la{\langle}
\def\ra{\rangle}
\def\beq{\begin{equation}}
\def\eeq{\end{equation}}
\def\be{\begin{eqnarray}}
\def\ee{\end{eqnarray}}

\def\k2av{\la k_T^2\ra}
\newcommand{\f}[2]{\frac{#1}{#2}}
\newcommand{\dd}{ {\textrm d}}
\begin{document}
\title{The Nuclear Modification Factor at Large Rapidities}

\author{G.G. Barnaf\"oldi\address[RMKI]{KFKI Research Institute for 
	Particle and Nuclear Physics, \\ 
        P.O. Box 49, Budapest H-1525, Hungary}\address[ELTE]{E\"otv\"os 
            University,\\
	P\'azm\'any P\'eter s\'et\'any 1/A, Budapest H-1117, Hungary},
        P. L\'evai\addressmark[RMKI],
        G. Papp\addressmark[ELTE],
        and
        G. Fai\addressmark[ELTE]\address[KSU]{Center for Nuclear Research, 
	Kent State University,\\
	Kent OH-44242, USA} \thanks{Supported by 
           Hungarian OTKA T043455, T047050, 
           MTA-NSF-OTKA OISE-0435701, U.S. DE-FG02-86ER40251, 
           Szent-Gy\"orgyi Scholarship and German DAAD exchange program.}
        }


\maketitle

\begin{abstract}
RHIC data on high-$p_T$ hadron production display strong suppression
in a wide rapidity region, indicating 
strong induced energy loss for both transversally and longitudinally
traveling partons. We investigate the interplay of energy loss and 
rapidity dependence in a perturbative QCD improved parton model, and 
estimate the opacity of the produced hot matter in $AuAu$ collisions
at energies $\sqrt{s}=200$ $A$GeV and $63$ $A$GeV at different
rapidity values. Direction-dependent suppression offers the 
possibility to study the geometry of the hot matter.
\end{abstract}


\section{Introduction}\label{intro}

In heavy-ion collisions at RHIC energies we expect the formation of
hot and dense matter in the central space-time region of the collision.
One can measure the density profile of the color charges in this 
deconfined region using induced energy loss of partons traversing the 
dense region in different directions. Data from central rapidity show
the transversally propagating partons and their large energy loss
(see $AuAu$ data from PHENIX~\cite{PHENIXauau05} and STAR~\cite{STAR}).
At large rapidities strong suppression has been seen 
by BRAHMS~\cite{BRAHMS05,BRAHMS} in pion and hadron
production. This finding offers the possibility to 
analyze energy loss for different momentum directions.

We apply a perturbative QCD improved parton model to describe hadron
production at high $p_T$~\cite{Yi02,bgg04}. 
The Glauber-Gribov model provides the proper framework to describe 
hadron production as a superposition of proton-proton results to
obtain baseline hadron spectra in heavy-ion collisions. 
The parton energy loss is calculated by the GLV-method~\cite{glv},
and the opacity of the dense matter is extracted in different rapidity
regions, i.e. for different parton momentum directions.
We include shadowing and multiscattering, tested in a wide rapidity 
range in $dAu$ collisions~\cite{dAu,Barnafqm04}.

\newpage
\section{Theoretical Background of the Model} 

\noindent 
Pion production is calculated in $AuAu$ collisions at RHIC energies
using a perturbative QCD improved parton model 
in leading order (LO)~\cite{Yi02,bgg04}. Schematically
\begin{equation}
\label{hadX}
E_{\pi} \f{\dd \sigma_{\pi}^{pA}}{ \dd ^3p} \sim 
    \sum\limits_{abc} T_{AA'}({\bf r},{\bf b}) \otimes 
    f_{a/p}(x_a,{\bf k}_{Ta},Q^2) \otimes
    f_{b/A}(x_b,{\bf k}_{Tb},Q^2) \otimes 
    \dd \hat{\sigma} \otimes D_{c}^{\pi} (z_c, \tilde{Q}^2)  \,\,  . 
\end{equation}
The collision geometry and the superposition of the nucleon-nucleon
collisions is included by the Glauber-Gribov model.
In the multidimensional convolution integral (\ref{hadX}),
$T_{AA'}({\bf r},{\bf b})=t_A(r) t_{A'}(|{\bf b} - {\bf r}|)$ is the
nuclear overlap function for the colliding nuclei.
The nuclear thickness function for nucleus $A$ is introduced as 
$t_{A}(b) = \int \dd z \, \rho_{A}(b,z)$, normalized as 
$\int \dd ^2b \, t_{A}(b)$ $= A$. We use   
a Woods\,--\,Saxon distribution for $Au$.

Nuclear PDFs can be constructed from proton and neutron PDFs~\cite{Yi02}:
\begin{equation}
f_{a/A}(x,Q^2) =
S_{a/A}(x) \left[ \,\, \frac{Z}{A} f_{a/p}(x,Q^2) + \left(1-\frac{Z}{A}\right)
  f_{a/n}(x,Q^2) \,\, \right]   \,\, . 
\label{shadow}
\end{equation}
The function $S_{a/A}(x)$ describes `conventional' nuclear 
shadowing. Here a $b$-independent $S_{a/A}(x)$ was taken from the HIJING 
parametrization~\cite{Shadxnw_uj}.

The last term in eq.~(\ref{hadX}) is
the fragmentation function (FF), $D_{c}^{\pi}(z_c, \tilde{Q}^2)$. This 
term gives
the probability for parton $c$ to fragment into a pion with momentum
fraction $z_c$ at scale $\tilde{Q}= \tilde{\kappa }\,  p_T$. 
We apply the KKP parametrization~\cite{KKP}.  
The factorization scale ($Q$) is connected 
to the momentum of the intermediate jet, $Q=\kappa \cdot p_T/z_c$, which 
scales together with $\tilde{Q} $ in the FF.
We applied $\kappa =\tilde{\kappa } = 2/3$ at RHIC energies.

The density of the color particles can be measured by the
induced energy loss of high-energy quarks and gluons traveling through
the hot dense deconfined matter. One can determine the  
non-abelian radiative energy loss, $\Delta E (E,L) $.
This quantity depends on the color charge density through the opacity,  
$\bar{n}=L/ \lambda $, where
$L$ is the length of the traversed matter and $\lambda$ is the mean
free path. In "thin plasma" approximation the
energy loss is given by the following expression~\cite{glv}:
\begin{equation}
\label{glv}
\Delta E_{GLV}
= \frac{C_R \alpha_s}{N(E)} \frac{L^2 \mu^2}{\lambda}
\log\left( \frac{E}{\mu} \right) \,\, .
\end{equation}
Here $C_R$ is the color Casimir of the jet  and
$\mu^2/\lambda \sim \alpha_s^2 \rho_{part}$
is a transport coefficient of the medium, which is
proportional to the parton density, $\rho_{part}$.
The color Debye screening scale is denoted by $\mu$.
$N(E)$ is an energy dependent numerical factor with 
an asymptotic value 4 at high jet energies.

Jet energy loss influences the final hadron spectra.
We can introduce this effect via modifying the 
momentum fraction of the outgoing parton before the fragmentation.
Considering an average energy loss, $\Delta E$, in a static plasma,
the argument of the fragmentation functions will be modified as
\begin{equation}
\label{quenchff}
\frac{D_{\pi/c} ( z_c , \tilde{Q}^2 )}{\pi z_c^2 } \ \ \longrightarrow
\ \ \frac{z^{\ast}_c}{z_c}  \,\,
\frac{D_{\pi/c} ( z^{\ast}_c , \tilde{Q}^2 )}{\pi z_c^2 } \,\, .
\end{equation}
Here $ z^{\ast}_c = z_c / \left(1- \Delta E/p_c \right) $
is the modified momentum fraction.

\section{Nuclear modification factor, $R^{\pi}_{AA}$, at large rapidities}

\begin{figure}[htb]
\begin{center}
\resizebox{1.0\textwidth}{!}{\includegraphics{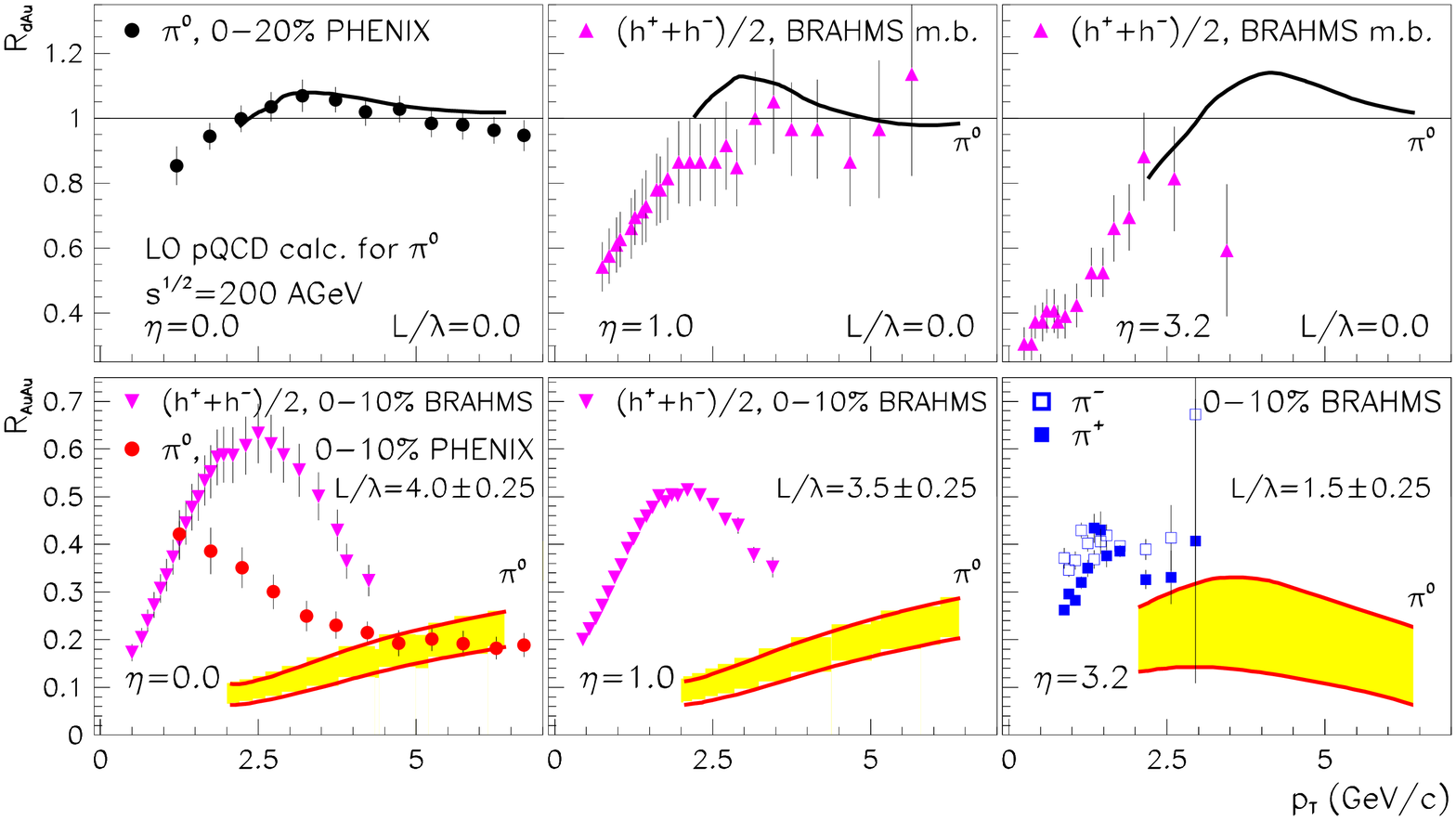}}
\end{center}
\vspace*{-1.2truecm}
\caption{
Nuclear modification factors at different rapidities: 
$R^{\pi}_{dAu} $ in $dAu$ collisions ({\sl upper row}),
$R^{\pi}_{AuAu}$ in $0-10\%$ central $AuAu$ collisions ({\sl lower row}).
Data are from 
PHENIX~\cite{PHENIXauau05,PHENIXdau05} and 
BRAHMS~\cite{BRAHMS05,BRAHMSdau05} at $\sqrt{s}=200 $ $A$GeV.
Full lines display our pQCD results, the bands indicate the
theoretical uncertainties on the value of the opacity.}
\label{fig1}
\end{figure}

Fig.~\ref{fig1}  displays the calculated nuclear modification factor 
for pion production in minimum bias $dAu$ ({\sl upper row})
and in central $AuAu$ ({\sl lower row}) collisions at $\sqrt{s}=200$~$A$GeV at 
different rapidities, $\eta=0,\, 1, \, 3.2$.  
The $dAu$ results indicate the validity of our LO perturbative QCD
calculations including multiscattering and shadowing~\cite{dAu,Barnafqm04}.
Data on pion and charge hadron production are shown from 
PHENIX~\cite{PHENIXdau05} and BRAHMS~\cite{BRAHMSdau05}.

\vspace*{-1.3truecm}

\vspace*{-0.5truecm}
In $AuAu$ collisions strong final state interaction (induced jet 
energy loss) appears and data display suppression at high 
$p_T$~\cite{PHENIXauau05}. At midrapidity, our pQCD result with average 
opacity of $L/\lambda= 4.0\pm 0.25$ is consistent with data 
at $p_T \geq 4$ GeV/c, 
where the high-$p_T$ region sets in (see Fig.~\ref{fig1}). 
Recent BRAHMS data~\cite{BRAHMS05,BRAHMS} may indicate 
rapidity independent suppression for pion production in the 
intermediate-$p_T$ region ($p_T < 3.5$~GeV/c).
Assuming that this finding is also valid for the high-$p_T$ region,
we wish to achieve a similar high-$p_T$ suppression at large rapidities.
A smaller opacity, $L/\lambda= 3.5\pm 0.25$ at $\eta=1$,
and $L/\lambda= 1.5\pm 0.25$ at $\eta=3.2$ is needed to do this.
(Note the difference between charge hadron and pion suppression.) 
This result indicates that longitudinally traveling partons see less 
colored matter then those traveling in the transverse direction.  

\begin{figure}[htb]
\vspace*{-0.8truecm}
\begin{center}
\resizebox{1.0\textwidth}{!}{\includegraphics{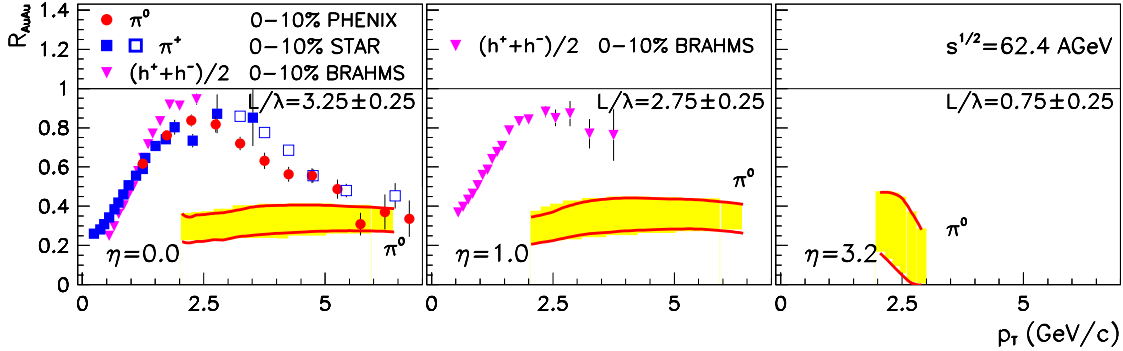}}
\end{center}
\vspace*{-1.1truecm}
\caption{Nuclear modification factor, 
$R^{\pi}_{AuAu}$,
at $\eta=0$, $1$ and $3.2$
in $0-10\%$ central $AuAu$ collisions at RHIC energy,
$\sqrt{s}=62.4 $ $A$GeV.
Our pQCD results with varying opacities (full lines and bands) are compared to 
PHENIX~\cite{PHENIXauau05}, STAR~\cite{STAR}, and BRAHMS~\cite{BRAHMS05} data.}
\label{fig2}
\end{figure}

Fig.~\ref{fig2} displays available data and our pQCD results for 
$AuAu$ collisions at $\sqrt{s} = 62.4$~$A$GeV for different $\eta$ values.
Due to the smaller Bjorken energy density~\cite{BRAHMS05,DEnterria},  
$\varepsilon_{Bj}^{62.4} < \varepsilon_{Bj}^{200}$,
we have used reduced opacity parameters, namely  
 $L/\lambda = 3.25\pm 0.25$ at mid-rapidity,
 $L/\lambda = 2.75\pm 0.25$ at $\eta=1$.
 and
 $L/\lambda = 0.75\pm 0.25$ at $\eta=3.2$.
Extended bands indicate the size of the theoretical uncertainties 
on the induced energy loss. 

\newpage

\vspace*{-0.5truecm}

\vspace*{-0.5truecm}
Comparing $R^{\pi}_{AuAu}$ results of Fig.~\ref{fig1} and 
Fig.~\ref{fig2}, the decreasing tendency of the opacity parameter is  
clearly seen as $\eta$ is increasing. 
At large forward rapidities the 
interplay between stronger shadowing and weaker quenching
effects maintains a rapidity independent
nuclear modification factor (no data are available at $\eta =3.2$ 
for $\sqrt{s} = 62.4$~$A$GeV $AuAu$).

Whether $\pi$ suppression is approximately rapidity independent will of
course be decided by the data. The present hint of this behavior
motivated us in this study to examine the interplay of rapidity dependence 
and jet energy loss. One particle quark tomography combined with
jet-jet correlation results provides information on the geometry 
of the hot region. Energy dependence and $CuCu$ results may be 
used to verify our conclusions. 



\begin{thebibliography}{99}  

\bibitem{PHENIXauau05}
    V. Greene for the [PHENIX Coll.], {\it these Proceedings}. 

\bibitem{STAR}
    J. Dunlop for the [STAR Coll.], {\it these Proceedings}. 

\bibitem{BRAHMS05}
    P. Staszel for the [BRAHMS Coll.], {\it these Proceedings}.


\bibitem{BRAHMS}
    D. R\"ohrich for the [BRAHMS Coll.], {\it these Proceedings}.

\bibitem{Yi02}
    Y. Zhang {\it et al.}, 
    {\it Phys. Rev.} {\bf C65} (2002) 034903.

\bibitem{bgg04}  
    G.G. Barnaf\"oldi {\it et al.},
    {\it Eur. Phys. J.} {\bf C33} (2004) S609.

\bibitem{glv}
    M. Gyulassy, P. L\'evai, I. Vitev,
    Phys. Rev. Lett. {\bf 85} (2000) 5535; 
    Nucl. Phys. {\bf B571} (2000) 197; 
    {\it ibid.} {\bf B594} (2001) 371.

\bibitem{dAu}
    P. L\'evai {\it et al.},
    {\tt nucl-th/0306019}.

\bibitem{Barnafqm04}  
    G.G. Barnaf\"oldi {\it et al.}, 
    {\it J. Phys.} {\bf G30} (2004) S1125.

\bibitem{Shadxnw_uj}
    S.J. Li and X.N. Wang, 
    {\it Phys. Lett.} {\bf B527} (2002) 85.

\bibitem{KKP}
    B.A. Kniehl, G. Kramer, and B. P{\"o}tter, 
    {\it Nucl. Phys.} {\bf B597} (2001) 337.

\bibitem{PHENIXdau05}
    B.A. Cole for the [PHENIX Coll.], {\it these Proceedings}. 


\bibitem{BRAHMSdau05}
   I. Arsene {\it et al.} [BRAHMS Coll.],
    {\it Phy. Rev. Lett.} {\bf 93} (2004) 242303; 
     Z. Yin {\it et al.} [BRAHMS Coll.], 
    {\it J. Phys.} {\bf G30} (2004) S983.

\bibitem{DEnterria}
    D. D'Enterria, {\it Phys. Letts.} {\bf B596} (2004) 32. 


\end{thebibliography}
\end{document}